\documentclass[useAMS,usenatbib,times,epsfig]{mn2e}
\usepackage{epsfig,times,amsmath, amsfonts, amssymb}

\title[Potential sources of contamination to weak lensing]{Potential sources of contamination to weak lensing measurements: constraints from N-body simulations} 
\author[Heymans et al.]{Catherine Heymans$^{1}$\thanks{heymans@physics.ubc.ca}, Martin White$^{2,3}$, Alan Heavens$^4$, Chris Vale$^{5,2}$ \& \newauthor
Ludovic Van Waerbeke$^1$\\
$^1$ Department of Physics and Astronomy, 6224 Agricultural Road, University of British Columbia, Vancouver, BC, V6T 1Z1, Canada.\\ 
$^2$ Department of Physics and Astronomy, 601 Campbell Hall, University of California Berkeley, CA 94720, USA.\\
$^3$ Lawrence Berkeley National Laboratory, 1 Cyclotron Road, Berkeley, CA 94720, USA.\\
$^4$ SUPA\thanks{Scottish Universities Physics Alliance}, Institute for Astronomy, University of Edinburgh, Blackford Hill, Edinburgh, EH9 3HJ, UK.\\
$^5$ Theoretical Astrophysics, Fermi National Accelerator Laboratory, Batavia, IL 60510, USA. \\ }

\newcommand{\be}{\begin{equation}}  \newcommand{\ee}{\end{equation}}
\newcommand{\bes}{\begin{equation*}}  \newcommand{\ees}{\end{equation*}}
  \newcommand{\ba}{\begin{eqnarray}}
\newcommand{\ea}{\end{eqnarray}}

\newcommand{\myplus}{+}
\newcommand{\myminus}{-}

\def\gs{\mathrel{\raise1.16pt\hbox{$>$}\kern-7.0pt 
\lower3.06pt\hbox{{$\scriptstyle \sim$}}}}         
\def\ls{\mathrel{\raise1.16pt\hbox{$<$}\kern-7.0pt 
\lower3.06pt\hbox{{$\scriptstyle \sim$}}}}

\begin{document}

\pagerange{\pageref{firstpage}--\pageref{lastpage}} \pubyear{2006}
\maketitle
\label{firstpage}

\begin{abstract}

We investigate the expected correlation between the weak gravitational
shear of distant galaxies and the orientation of foreground galaxies,
through the use of numerical simulations.  This shear-ellipticity
correlation can mimic a cosmological weak lensing signal, and is
potentially the limiting physical systematic effect for cosmology with
future high-precision weak lensing surveys.  We find that, if
uncorrected, the shear-ellipticity correlation could contribute up to
10\% of the weak lensing signal on scales up to 20 arcminutes, for
lensing surveys with a median depth $z_m=1$.  The most massive foreground
galaxies are expected to cause the largest 
correlations, a result also seen in
the Sloan Digital Sky Survey.  We find that the redshift dependence of
the effect is proportional to the lensing efficiency of the
foreground, and this offers prospects for removal to high precision,
although with some model dependence. The contamination is
characterised by a weakly negative B-mode, which can be used as a diagnostic
of systematic errors.  We also provide more accurate predictions for a
second potential source of error, the intrinsic alignment of nearby galaxies. 
This source of contamination is less important, however, 
as it can be easily removed with distance information.

\end{abstract}


\section{Introduction}

Weak gravitational lensing by large-scale structure provides a test of
cosmology that is based on the well understood physics of gravity. 
The lensing strength is directly related to the large-scale distribution of
matter and thus enables constraints to be placed on cosmological
parameters such as the matter density 
parameter $\Omega_{\rm m}$ and the amplitude of the 
matter power spectrum $\sigma_8$\citep{vWb01,HYG02,BMRE,Jarvis,MLB02,Hamana,Massey,RhodesSTIS,vWb04,HymzGEMS,CFHTLS_deep,CFHTLS_wide}.
Combined with redshift information, measurements of weak 
lensing are also able to constrain the equation of state of the dark energy
$w(z)$ as lensing can probe 
the effect dark energy has on both the evolution of the matter 
power spectrum and the redshift-distance relationship.  
With the next generation of deep-field and 
wide-field multi-colour surveys, weak lensing has the exciting prospect 
of providing constraints to distinguish between a cosmological constant
 ($w=-1$) and quintessence models ($w(z) \ge -1$).
 
%Cosmological shear is weak, at the percent level, and hence
%the signal is technically very challenging to measure.  To address this, weak
%lensing methods have been under constant review and development such that weak
%lensing studies have reached a stage where measurements need not be limited by
%the systematic errors introduced by the non-lensing distortions in survey
%observations\citep{STEP}.   

Lensing by large-scale structure weakly distorts
images of background galaxies such that the observed galaxy ellipticity 
\be
e_{\rm obs} \simeq e_{\rm s} + \gamma \,\, ,
\label{eqn:ellishear}
\ee
where $e_{\rm s}$ is the intrinsic galaxy ellipticity, or source ellipticity, 
and $\gamma$ is the weak lensing shear distortion.
Lensing by large-scale structure is coherent
and hence it induces weak correlations in the 
observed ellipticities of galaxies such that 
$\langle e^a_{\rm obs}e^b_{\rm obs}\rangle _\theta > 0$, where the average is taken
over all galaxy pairs $(a,b)$ separated by angle $\theta$.  
From equation~\ref{eqn:ellishear},
\be
\langle e^a_{\rm obs}e^b_{\rm obs}\rangle  = \langle e^a_{\rm s} e^b_{\rm s}\rangle  + \langle \gamma^a e^b_{\rm s}\rangle  + \langle \gamma^b e^a_{\rm s}\rangle  + \langle \gamma^a \gamma^b\rangle  \,\, .
\label{eqn:ellicor}
\ee
In order to extract the weak cosmological signal $\langle \gamma^a \gamma^b\rangle $ 
from the observed galaxy ellipticity correlation 
$\langle e^a_{\rm obs}e^b_{\rm obs}\rangle $, all intrinsic terms including $e_{\rm s}$ 
are typically assumed to be negligible.   
There are however potential physical limitations to this assumption
related to correlations
between galaxy shape and the density field. This could arise, for example,
through tidal torques operating during
galaxy formation \citep{Hoyle,P69,HP88}.  Such a
correlation would introduce an intrinsic alignment between 
nearby galaxies mimicking the weak lensing effect, i.e if galaxies $a$ and $b$
were physically close $\langle e^a_{\rm s} e^b_{\rm s}\rangle  \ne 0$.  Investigated
observationally (\citet{BTHD02,LP02,HBH04}, hereafter H04, \citet{RM05}), 
analytically \citep{CKB01,CNPT01,LP01,Porciani,Mackey02} 
and through numerical simulations (\citet{HRH00}, hereafter
HRH, \citet{CM00,Jing}, H04), it is 
currently thought that this intrinsic alignment
effect is weak, at the level of a few percent contamination
to the lensing signal from a deep survey of median redshift 
$z_m \sim 1$. Division
of a redshift survey for lensing tomography \citep[see for example][]{Hu99}, 
however, 
significantly increases the contamination in the lower redshift bins (H04).  As
shown by \citet{HH03} and \citet{KingSch02} this contamination can be
effectively 
removed at high accuracy 
by optimally downweighting pairs of nearby galaxies in the lensing
analysis (see H04 for an implementation of
this method).  Hence this physical systematic need not limit the accuracy of
future multi-colour weak lensing surveys.  

Correlations between a galaxy's shape and its surrounding 
density field may also introduce
another more subtle effect, first discussed by \citet{HS04}.
Consider the image of a distant galaxy lensed by the dark matter density 
surrounding a foreground galaxy.  If the shape of the foreground
galaxy is correlated with the tidal gravitational field, 
this could lead to a non-zero
correlation\footnote{We hereafter refer to this effect as a
shear-ellipticity correlation.  In the literature \citet{HS04} and 
\citet{King05} refer to this 
effect as intrinsic alignment-lensing interference and \citet{RM05} use the 
term GI contamination.} between intrinsic foreground galaxy ellipticity 
$e_{\rm s}$ and the 
lensing shear distortion experienced by the distant galaxy $\gamma$, i.e 
if galaxy $a$ is always more distant than galaxy $b$, 
$\langle \gamma^a e^b_{\rm s}\rangle  \ne 0$.
\citet{HS04} explore two analytical models to estimate the
magnitude of this effect finding that when 
galaxy ellipticity is quadratically related to the tidal or density 
field, as suggested by tidal torque theory, no shear-ellipticity
correlation is found.  When galaxy ellipticity is linearly
related to the tidal or density field, however, 
the shear-ellipticity correlation can
be similar to the magnitude of the weak lensing shear correlation.  
In this model, galaxies are assumed initially to be spherical and 
acquire ellipticity by responding to the linear tidal field.
For a spherically symmetric over-density, the tidal field compresses the 
spherical galaxy along the direction to the over-density centre.  
Around a spherically symmetric proto-void, the linear theory would stretch the
galaxy along the direction to the void centre.  
There is some observational evidence to support the linear model from 
\citet{Trujillo}, who find that the 
rotation axes of galaxies located near voids are preferentially aligned 
with the void surface.  
%The traditional models of galaxy formation would 
%however favour the quadratic model.

If a shear-ellipticity correlation does exist 
then its strength is expected to be 
dependent on the lensing efficiency of the lens-source pair \citep{HS04}.  
\citet{King05} has developed this idea to show that using correlation 
function tomography and knowledge of the redshift dependence of each 
source of contamination, one could separate the true cosmological 
lensing signal 
from both the intrinsic galaxy alignment contamination and the
shear-ellipticity correlation contamination.  

The shear-ellipticity correlation 
effect has recently been investigated observationally by
\citet{RM05} who measure the correlation between the intrinsic
ellipticities of nearby galaxies and the density field that surrounds
them in the Sloan Digital Sky Survey (SDSS).  
As the density field is responsible for the
gravitational lensing of background galaxies, \citet{RM05} are able to
estimate the shear-ellipticity correlation for any lensing survey. 
They find that for most galaxy types, the shear-ellipticity correlation 
is consistent with no signal.  The shapes of the most luminous
red galaxies are however found to be be strongly correlated with the 
density field.  For a lensing survey with
a median source redshift $z_m \sim 1$
the most luminous galaxies would then introduce a 
shear-ellipticity correlation signal 
that is found to be at a similar magnitude
to the pure lensing correlation signal from a
$\Lambda$CDM cosmology.  Averaging over the full galaxy sample, \citet{RM05} 
suggest that there would be a 
$\sim 5\%$ contamination to the pure lensing signal.  These  
results are however subject to noise 
(the results for the full galaxy sample are also consistent with zero contamination), and are based on measurements taken at very low 
redshift and projected to the higher redshifts that are 
probed by lensing surveys.  In this paper we therefore aim to address this 
issue through the analysis of extensive 
numerical dark matter simulations \citep{ValeWhite},
following the philosophy of HRH and H04 whereby galaxy ellipticity is
defined from the properties of the parent dark matter halos. 

This paper is organised as follows.  In section~\ref{sec:sims} we
describe the N-body simulations that we use in this analysis and the
different models used to populate dark matter halos with luminous
galaxies.   We update the results of HRH and H04 in
section~\ref{sec:IAres} by presenting higher signal-to-noise
measurements of the E and B-mode intrinsic alignment correlation
$\langle e^a_{\rm s} e^b_{\rm s}\rangle _\theta$ from our 
N-body simulations and we
compare these results to the recent observational measurements of
\citet{RM05}.  We also investigate the effect of nearby voids on galaxy 
shape and compare our results
to the observational measurements of \citet{Trujillo}.
In section~\ref{sec:res} we present measurements of the
E and B-mode shear-ellipticity correlation 
$\langle \gamma^a e^b_{\rm s}\rangle $ 
and measure the redshift dependence of this signal.  This allows us to 
provide fitting formulae for the shear-ellipticity correlation measured 
from different galaxy models.
We discuss our findings in section~\ref{sec:discussion} 
and conclude in section~\ref{sec:conc}.

\section{N-body simulations}
\label{sec:sims}

The basis of our calculation is a large N-body simulation of a $\Lambda$CDM
cosmology.  The simulation used $512^3$ particles in a periodic cubic box
$300h^{-1}$Mpc on a side.
This represents a large enough cosmological volume to ensure a fair sample
of the Universe, while maintaining enough mass resolution to identify
galactic mass halos.
The cosmological model is chosen to provide a reasonable fit to a wide range
of observations with $\Omega_{\rm m}=0.3$, $\Omega_\Lambda=0.7$,
$H_0=100\,h\,{\rm km}{\rm s}^{-1}{\rm Mpc}^{-1}$ with $h=0.7$,
$\Omega_{\rm B}h^2=0.02$ and a scale-invariant spectrum with $\sigma_8=0.8$.
The simulation was started at $z=60$ and evolved to $z=0$ using the
TreePM code of \citet{TreePM}.  The full phase space distribution was recorded
every $100h^{-1}$Mpc from $z\simeq 2$ to $z=0$.
The gravitational force softening was of a spline form, with a
``Plummer-equivalent'' softening length of $20\,h^{-1}$kpc comoving.
The particle mass is $1.7\times 10^{10}h^{-1}{\rm M}_\odot$ allowing us to find
bound halos with masses several times $10^{11}h^{-1}{\rm M}_\odot$.

For each output we determine the halo catalogue by running a
``friends-of-friends'' (FoF) group finder (e.g.~\cite{DEFW}) with
a linking length $b=0.15$ in units of the mean inter-particle spacing.
This procedure partitions the particles into equivalence classes, by
linking together all particle pairs separated by less than a distance $b$.
This means that FoF halos are bounded by a surface of density roughly $140$
times the background density.  For each halo we computed the angular momentum
and the moment of inertia tensor, using the centre of mass of
the particles in the FoF group as the centre of the halo.

A past light cone was constructed by stacking the boxes back along the line
of sight, with boxes shifted by random amounts in the $x$- and $y$-directions
and viewed down randomly chosen axes to avoid artificial periodicity.  The
full projected mass distribution was used to create convergence maps of 
$\kappa$ at a source redshift plane of $z_s \sim 1$ and $z_s \sim 0.5$, where
$\kappa$ is defined to be the dimensionless surface mass density. 
The gravitational shear $\gamma = (\gamma_1, \gamma_2)$ 
that a galaxy at $z_s \sim 1$, or $z_s \sim 0.5$ would experience is then
created from the $\kappa$ maps \citep{KS93}.  
The halo
information was transformed into the same coordinate system to create a light
cone halo distribution.  $12$ different realizations of the $2048^2$ pixel
maps were made in all, each with a 25 square degree field-of-view.

\subsection{Galaxy Models}
\label{sec:galmods}

From a catalogue of dark matter halos, 
there are two steps required to simulate luminous
galaxies with ellipticities.  Firstly we need to populate the halos with 
galaxies, for which, in this paper, we use the 
conditional luminosity function (CLF) of \citet{CLF_CM05}, 
and secondly we need to assign ellipticities.  
We follow HRH and H04 in relating the ellipticities $e$ of the galaxies to the 
ellipticity of the halo (the `elliptical' model) or via the angular 
momentum of the halo (the `spiral' model). 
In this section we first describe the CLF 
halo occupation model, and then discuss the elliptical, spiral and galaxy mix 
models that we use in the analysis that follows.

The CLF, denoted by $\Phi(L|M_h,z)$ gives the
average number of galaxies with luminosities between $L$ and $L+dL$
that reside in halos of mass $M_h$ at a redshift of $z$.  The CLF model
is separated into terms associated with central and satellite galaxies
with a further division into early-type and late-type galaxies.  The free
model parameters are then constrained in \citet{Cooray06} 
by observations of the galaxy
luminosity function and galaxy correlation function over a wide range
of redshift samples (for example the SDSS at
$z<0.1$ \citep{zehavi04} and the COMBO-17 survey at $0.4<z<0.8$
\citep{Wolf03}).  The results of this analysis show a very
convincing match between CLF model predictions and observational
results at all redshifts and thus provide an adequate model with which
to populate simulated dark matter halos with luminous galaxies.  
The average
number of galaxies that reside in a halo of mass $M_h$ at redshift $z$
is given by $N(M_h,z) = \int \,dL\, \Phi(L|M_h,z)$ where the luminosity ranges
from the faintest observable magnitude (which we set to a typical
ground-based survey depth of $r<25.5$), to an absolute brightest galaxy
magnitude $M_r > - 23$.  We use the best fit \citet{Cooray06} 
CLF parameters or the chosen
fiducial values in the cases where the parameters are poorly constrained.
This results in a realistic redshift distribution of the foreground 
galaxies with roughly 15 galaxies per square arcmin with $z<1.0$.

We use three simple galaxy models which we term, ellipticals, spirals
and mix.  Ellipticals are defined to have the same ellipticity as
their parent halos.  Spirals are modelled as a thick disk randomly
misaligned with the angular momentum vector is its parent halo, with a
mean misalignment angle of $20^\circ$ as suggested by the results of
\citet{vdBosch02} and discussed in more detail in H04.  Note that for 
completeness, we also present the results, in section~\ref{sec:IAres}, 
for a spiral model 
where the disk is aligned with the angular momentum vector of its parent halo, 
as in HRH.  The results in the shear-ellipticity correlation analysis that 
follow in section~\ref{sec:res}, 
however, are unaffected by whether the disk is 
misaligned or not, and we therefore only present the
shear-ellipticity correlation results from our 
misaligned spiral model.  Our mix model
populates dark matter halos with the same elliptical or misaligned spiral
galaxies according to the early-type and late-type CLFs of \citet{Cooray06}.
These CLFs are constrained such that they reproduce 
the luminosity functions and clustering properties of red and blue galaxies. 
Our simple galaxy ellipticity models will clearly break down in the
case of parent halos which host more than one galaxy.  We therefore
limit our study to singly occupied halos which satellite galaxies are
unlikely to inhabit. For our mix model, this criterion results with
roughly one third of the halos populated with elliptical galaxies at all 
redshifts.  

In this paper we determine the 
ellipticity parameters $\varepsilon_i$ for each galaxy, where
$\varepsilon_i$ for a perfect ellipse with axial ratio $\beta$ at
position angle $\theta$, is given by,
\be
\left(
\begin{array}{c}
\varepsilon_1 \\
\varepsilon_2
\end{array}
\left)
= \frac{1-\beta^2}{1+\beta^2}
\right(
\begin{array}{c}
\cos 2\theta \\
\sin 2 \theta
\end{array}
\right)\, .
\label{eqn:elliparam}
\ee
In order to study the effect of these intrinsic galaxy shapes on weak lensing measurements we then define and use the alternative ellipticity parameters $e_i$ 
given by
\be 
e_i = \frac{\varepsilon_i}{2 {\it R}} \,\, ,
\label{eqn:resp}
\ee
where the responsivity ${\it R}$ is a correction to account for the 
response of an ensemble of galaxies to weak lensing shear 
\citep{RRG00,Bernstein}.
Without this responsivity correction to our ellipticity measurements, 
equation~\ref{eqn:ellishear} would not apply.
For the following unweighted analysis,  
${\it R} = 1 - \langle \varepsilon_i^2\rangle  = 0.89$ for all galaxy models 
(although note that for simplicity in HRH and H04 
the responsivity ${\it R}$ was set equal to 1).  The 
resulting $e_i$ ellipticity distributions have a mean 
$\langle e_i \rangle = 0$ and a width $\sigma_e = 0.33$ which is very 
similar to what is measured from space-based data \citep{GEMSgg}.

\section{Intrinsic Galaxy Alignments}
\label{sec:IAres}
\begin{figure}
\begin{center}
\epsfig{file=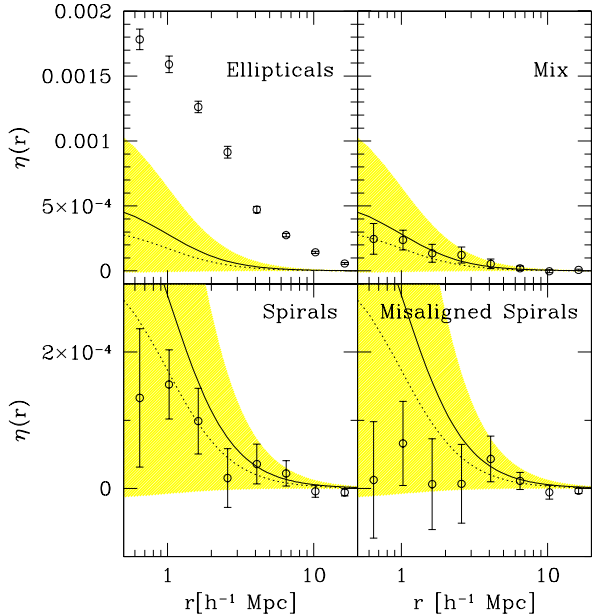,width=8.5cm,angle=0,clip=}
\caption{The intrinsic galaxy ellipticity correlation $\eta(r_{ab})$ as a 
function of comoving galaxy separation $r_{ab}$ as measured from the 
simulation assuming three different galaxy models; ellipticals, 
spirals (placed perpendicular to the angular momentum vector as in HRH, 
lower left, and misaligned, as in H04 and the rest of this paper, lower right) 
and mix. 
The shaded region shows the $95\%$ confidence 
region constrained from measurements of intrinsic galaxy alignments 
in the SDSS \citet{RM05}, where the best-fit is over-plotted (solid). 
The best fit model from the misaligned spiral galaxies measured in H04 
is shown dotted.}
\label{fig:IA_nr}
\end{center}
\end{figure}

In this section we update the analysis of HRH and H04 by re-measuring
the intrinsic alignment correlation 
$\langle e^a_{\rm s} e^b_{\rm s}\rangle $ from 
simulations.  The N-body
simulation used in this analysis is at a significantly higher
resolution than the VIRGO simulations \citep{Jenkins} used by HRH 
allowing for more reliable measurements of halo angular momentum.  
It also covers a larger cosmological volume and thus yields higher
signal-to-noise measurements.  The evolution of the large-scale
structure is now preserved, permitting a direct measurement of the 
E and B modes of the 
angular intrinsic alignment correlation.  As a result we do not have
to rely on projecting a measured real space correlation function
into an angular correlation function
using weakly constrained redshift dependent galaxy
clustering models.  The analyses of the simulations differ slightly
in the fact that we now populate halos according to the CLF of
\citet{Cooray06} instead of introducing low and high mass cuts to 
remove halos less likely to harbour a single  galaxy.  Our galaxy models 
are however identical, with the new addition of the mix model.
The intrinsic ellipticity correlation $\eta(r_{ab})$, is given by 
\be
\eta(r_{ab}) = \langle e_t({\bf
r_a}) e_t({\bf r_b})\rangle + \langle e_r({\bf
r_a}) e_r({\bf r_b})\rangle  \,\, , 
\ee
where the comoving galaxy separation
$r_{ab} = |{\bf r_a} - {\bf r_b}|$, and $\langle e_t e_t \rangle$ and 
$\langle e_r e_r \rangle$ are the tangential and rotated correlation 
functions (see for example H04).  The measured $\eta(r_{ab})$ 
is shown in Figure~\ref{fig:IA_nr}
for each galaxy model, where for completeness we have also included the 
HRH spiral model where the disk is placed perpendicular to the angular 
momentum vector of the halo.  The shaded region in each plot is the
$1\sigma$ confidence region for $\eta(r_{ab})$ as constrained from
measurements of intrinsic galaxy alignments 
in the SDSS \citep{RM05}.   
The strength of the signal for the elliptical model at
separations $r> 3 h^{-1} {\rm Mpc}$ is inconsistent with the
SDSS results at the $6\sigma$ level and we can thus reject this model
at $99.9\%$ confidence.  The spiral and mix models are consistent with
the SDSS constraints.  All results are consistent with 
the previous results from HRH and H04, and now supersede those constraints.

Following H04, we provide parameter fits to these results using the following 
model
\be
\langle e_i({\bf r_a}) e_i({\bf r_b})\rangle = \frac{A}{1 + (r_{ab}/B)^2} \,\, ,
\label{eqn:model}
\ee
where the amplitude $A$ is chosen to be a free parameter.  
For the elliptical model, $B$ is also allowed to be a free parameter.  
For all other models $B$ constrained to $B = 1\, h^{-1} {\rm Mpc}$.
Table~\ref{tab:fits} compares the best fitting parameters for the tangential 
$\langle e_t e_t \rangle$ and
rotated $\langle e_r e_r \rangle$ and
rotated $\langle e_r e_r \rangle$ ellipticity 
correlation functions, as well as the total 
ellipticity correlation $\eta(r_{ab})$.  For the elliptical model, 
both $A$ and $B$ are free parameters.  For all other models $B$ 
is constrained to be $B = 1 \, h^{-1} {\rm Mpc}$.
The best fits to the results 
presented in this paper can be compared to the result from H04 
(referred to as HRH* in that paper) and 
\citet{RM05} (SDSS), where we have included a scaling factor of $4R^2$ 
to account for the shear responsivity 
(equation~\ref{eqn:resp}) that we include in this analysis. 
It is interesting to note that the rotated component of the ellipticity 
correlation function $\langle e_r e_r \rangle$, is found to be weaker than 
the tangential component $\langle e_t e_t \rangle$, which is in contrast to 
previous assumptions that 
$\langle e_t e_t \rangle \simeq \langle e_r e_r \rangle$ .  
 
\begin{table}
\begin{center}
\begin{tabular}{l|c|c|c|c}
Model & Parameter & $\langle e_t e_t \rangle$ & $\langle e_r e_r \rangle$ & 
$\eta(r_{ab})$ \\ \hline\hline
Ellipticals   & $ A/10^{-3}$ & $   1.47^{+   0.05}_{-   0.02}$ & $   0.38^{+   0.18}_{-   0.08}$ & $   1.75^{+   0.08}_{-   0.02}$ \\ \hline
& $ B$ & $   2.94^{+   0.07}_{-   0.04}$ & $   0.96^{+   0.40}_{-   0.30}$ & $   2.77^{+   0.09}_{-   0.07}$ \\ \hline
Mix           & $ A/10^{-3}$ & $   0.45^{+   0.10}_{-   0.05}$ & $   0.07^{+   
.08}_{-   0.05}$ & $   0.47^{+   0.13}_{-   0.07}$ \\ \hline
Spirals       & $ A/10^{-3}$ & $   0.30^{+   0.05}_{-   0.02}$ & $  -0.02^{+   0.07}_{-   0.05}$ & $   0.28^{+   0.10}_{-   0.05}$ \\ \hline
Misaligned    & $ A/10^{-3}$ & $   0.15^{+   0.08}_{-   0.02}$ & $   0.00^{+   0.05}_{-   0.02}$ & $   0.07^{+   0.10}_{-   0.05}$ \\ \hline
HRH* & $ A/10^{-3}$ & - & - & $0.35 \pm 0.16$ \\\hline
SDSS & $ A/10^{-3}$ & - & - & $0.57 \pm 0.72$ \\
\end{tabular}
\end{center}
\caption{The intrinsic alignment amplitude $A$ and scale length $B$ as 
defined in equation~\ref{eqn:model} for the tangential 
$\langle e_t e_t \rangle$, and
rotated $\langle e_r e_r \rangle$ ellipticity 
correlation functions, as well as the total 
ellipticity correlation $\eta(r_{ab})$.  For the elliptical model, 
both $A$ and $B$ are free parameters.  For all other models $B$ 
is constrained to be $B = 1 \, h^{-1} {\rm Mpc}$.
The best fits to the results 
presented in this paper can be compared to the result from H04 (HRH*) and 
\citet{RM05} (SDSS), where we have included a scaling factor of $4R^2$ 
to account for the shear responsivity 
(equation~\ref{eqn:resp}) that we include in this analysis. 
The quoted errors are all $1\sigma$, except in the case of
SDSS which corresponds to a $3\sigma$ constraint.}
\label{tab:fits}
\end{table}

Taking all halos between a redshift of $0<z<1$ we measure the
angular galaxy ellipticity correlation 
$\langle e_{\rm s}({\bf \theta_a}) e_{\rm
s}({\bf \theta_b})\rangle $, directly from the simulations.
\citet{CNPT02} show that the correlation functions can be decomposed
into the following E- and B-type correlators, 
\be
\xi^E(\theta)=\frac{\xi_\myplus(\theta)+\xi'(\theta)}{2}\ \ \ \ \ \ 
\xi^B(\theta)=\frac{\xi_\myplus(\theta)-\xi'(\theta)}{2}
\label{eqn:xieb}
\ee
where, in this case,
\be 
\xi_{\pm}(\theta) = \langle e_t e_t \rangle_{\theta} 
\pm \langle e_r e_r \rangle_{\theta} ,
\label{eqn:xiplusminus}
\ee
and
\be
\xi'(\theta)=\xi_\myminus(\theta)+4\int_\theta^\infty \frac{d\vartheta}{\vartheta} \xi_\myminus(\vartheta)
        -12\theta^2 \int_\theta^\infty \frac{d\vartheta}{\vartheta^3}\xi_\myminus(\vartheta).
\label{eqn:xipr}
\ee

\begin{figure}
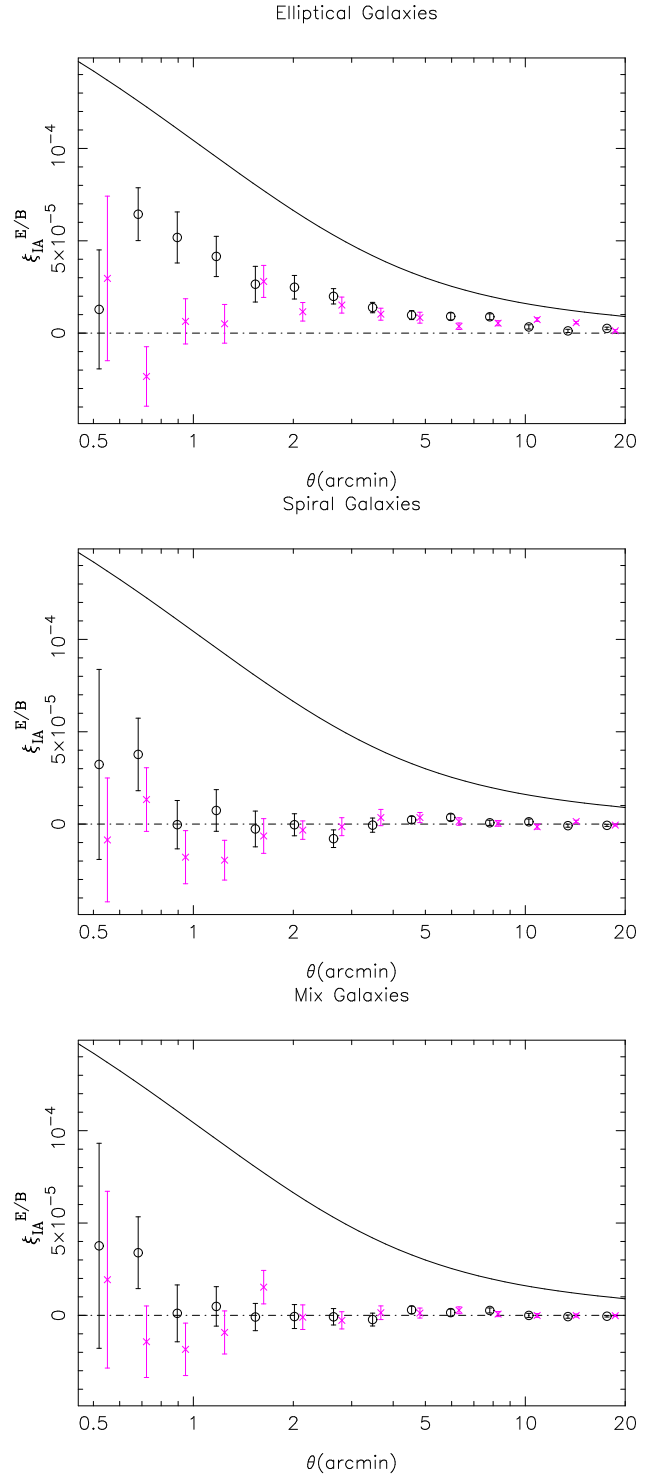

\begin{center}
\epsfig{file=IA_EB_eli.ps,width=6.5cm,angle=270,clip=}
\epsfig{file=IA_EB_sma.ps,width=6.5cm,angle=270,clip=}
\epsfig{file=IA_EB_mix.ps,width=6.5cm,angle=270,clip=}
\end{center}
\caption{The E mode (circle) and B mode (cross) 
decomposition of the angular intrinsic
galaxy ellipticity correlation $\langle e_s^a e_s^b\rangle _\theta$ as a  
function of
angular galaxy separation on the sky, $\theta$.  This is measured from the
simulation assuming three different galaxy models; ellipticals (upper panel),
spirals (middle panel) and mix (lower panel). 
The resulting E-mode can be compared to the expected
shear correlation (solid line) from a weak lensing survey with median
redshift $z_m = 0.7$, which corresponds to the median redshift of the 
galaxies in the simulations.  
For clarity the B-mode points are slightly offset to the right.}
\label{fig:IA_theta}
\end{figure}

\begin{figure}

\begin{center}
\epsfig{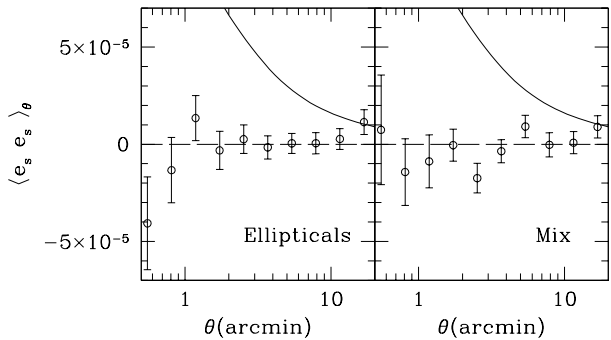}
\end{center}
\caption{The measured angular intrinsic
galaxy ellipticity correlation $\langle e_s^a e_s^b\rangle _\theta$ as a  
function of
angular galaxy separation on the sky, $\theta$ when galaxy 
pairs that have a physical separation of $r_{ab}<10 h^{-1} {\rm Mpc}$ 
are removed from the analysis.  This is measured from the
simulation assuming the elliptical galaxy model (right) and mix galaxy model 
(left).
The shear correlation from a weak lensing survey with median
redshift $z_m = 0.7$, which corresponds to the median redshift of the 
galaxies in the simulations, is shown for reference (solid line).}
\label{fig:IA_cp_rem}
\end{figure}

Following the method detailed in \citet{PenWM}, 
we have decomposed the ellipticity
correlation functions into their E and B modes $\xi^{E/B}$, 
assuming that $\langle e^a
e^b\rangle _\theta = 0$ for $\theta>20$ arcmin. 
Note that we treat each realization of the N-body simulations as an
independent pointing and plot the mean and co-variance between the
results from each realisation. The results are shown
in Figure~\ref{fig:IA_theta} where the measured E-mode
can be compared to the expected $\Lambda$CDM 
shear correlation from a weak lensing survey with median redshift 
$z_m \sim 0.7$.  This redshift
corresponds to the median redshift of the galaxies,
or the populated halos, 
in the simulations.  For both the spiral and mix models, 
the signal is consistent with zero on all scales, although 
see H04 for a noise-free
E/B mode analysis of the expected angular correlation signal from the
best-fitting model to $\eta(r_{ab})$, which is directly applicable to 
the spiral and mix models analysed in this work.
The elliptical model exhibits strong E and B modes 
out to $\theta < 10$ arcmin, although the small angle B-mode are 
consistent with zero.  Figure~\ref{fig:IA_cp_rem} shows the E/B mode 
decomposition of the elliptical model signal
when we have repeated the analysis removing galaxy 
pairs that have a physical separation of $r_{ab}<10 h^{-1} {\rm Mpc}$, 
the resulting E and B mode signal is consistent with zero on all scales, 
as expected.   


\subsection{Intrinsic Galaxy Alignment with Voids}
\label{sec:voids}
\begin{figure}
\begin{center}
\epsfig{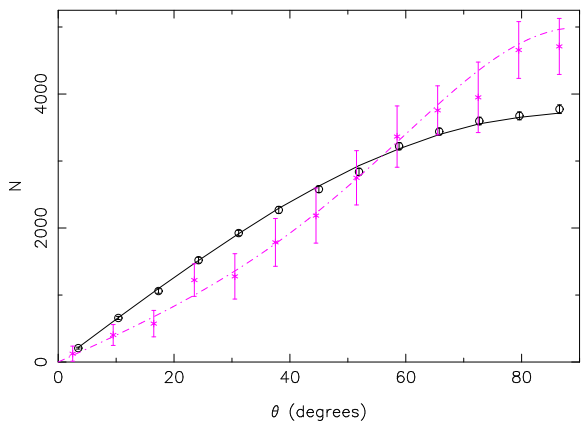}
\end{center}
\caption{The distribution of galaxies residing near to a void
as a function of their degree of alignment between the rotation axis
of the galaxies and the surface of the void (circles).  
For a perfectly aligned
set of galaxies we would expect to see a peak at $\theta=90^\circ$.
Instead our result is consistent with the null hypothesis that
galaxies are randomly oriented with all values of $\cos(\theta)$ equally 
likely.  The errors come from a bootstrap analysis, and the solid line is 
the average result when the halo angular momenta are randomised.  
Our results can be compared with the observational
results of \citet{Trujillo} (crosses) 
and the best fitting analytical prediction to these results from 
\citet{Lee04}, shown dot-dashed.}
\label{fig:voids}
\end{figure}

We follow the analysis of \citet{Trujillo} to determine the degree of
alignment between the rotation axis of galaxies located on the shells
of cosmic voids, and the surface of those voids.  We use a section of
the simulation at fixed output time $z \sim 0.325$ and take all halos
with mass $M_h> 8.5\times 10^{11}h^{-1}{\rm M}_\odot$ yielding a similar
number density of the 2dFGRS and SDSS surveys analysed by
\citet{Trujillo}.  Voids are selected using the same method of
\citet{Patiri} and for each galaxy within $r_{\rm void} < r < r_{\rm
void} + 4 h^{-1} {\rm Mpc}$ of the void centre (which has radius
$r_{\rm void}$) we calculate $\cos(\theta)={\bf r.L}/(|{\bf r}| |{\bf
L}|)$ where ${\bf L}$ is the angular momentum of the parent halo.
Figure~\ref{fig:voids} shows the resulting distribution of
galaxies residing on the surface of voids with $0 \le \theta \le 90$ degrees.  
The errors come from a bootstrap analysis, and the solid line is the average 
result when the halo angular momenta are randomised.
We find a distribution that is consistent with the null hypothesis
that the angular momentum vector of halos near voids is randomly
oriented with respect to the centre of voids.  For our spiral galaxy 
model, where galaxy orientation is determined by the angular momenta of the
parent dark matter halo, this result implies that no significant correlation 
exists between the orientation of galaxies 
and the centre of nearby voids in our simulations.
This is in contrast to the 
\citet{Trujillo} detection results, which are 
over-plotted on figure~\ref{fig:voids} 
that can reject this null hypothesis at the 
99.7\% confidence level.  These results are well fitted by the 
analytic prediction of \citet{Lee04}, (shown dot-dashed).
We discuss this difference further in 
section~\ref{sec:conc}.

\section{Shear-ellipticity correlations}
\label{sec:res}

\begin{figure}
\begin{center}
\epsfig{file=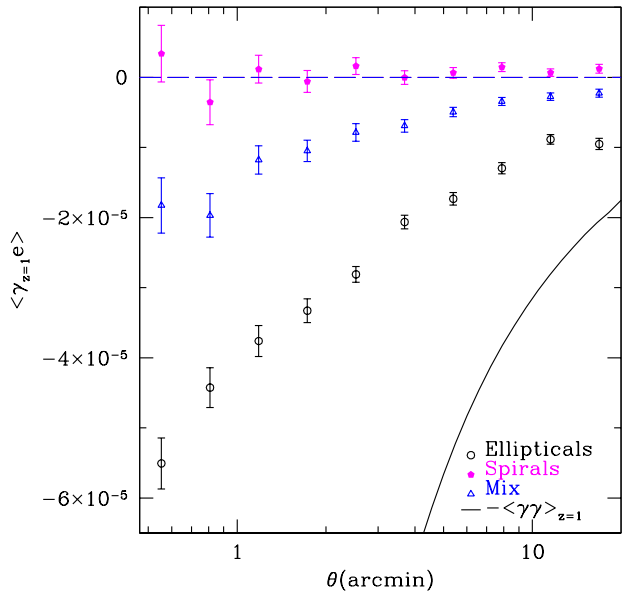,width=8.5cm,angle=0,clip=}
\end{center}
\caption{The shear-ellipticity correlation function $\langle \gamma e\rangle _\theta$ for the elliptical (circles), spiral (triangle) and mix (filled) galaxy models.  In this result all source galaxies (that experience a lensing 
shear $\gamma$) are placed at a source redshift $z_s = 1$.  
The foreground galaxies, 
with intrinsic ellipticity $e$, populate the redshift space $0<z<1$. 
These results are tabulated in table~\ref{tab:contam} 
along with the results for a 
survey were all source galaxies are placed at $z_s = 0.5$.
For reference, the solid theory curve shows 
the shear-shear correlation function (multiplied by $-1$) for a deep $z_s = 1$
survey using the adopted $\Lambda$CDM cosmology.}
\label{fig:shearecor}
\end{figure}

In this section we present shear-ellipticity correlation functions and
their E/B mode decomposition for the three different galaxy models, as
described in section~\ref{sec:galmods}.  Figure~\ref{fig:shearecor} shows
the measured correlation $\langle \gamma e \rangle$
between foreground galaxy ellipticity $e$ 
(where the foreground galaxies have redshifts $0<z<1$) and
the gravitational
shear $\gamma$ experienced by a set of background source galaxies
at $z_s \sim 1$ where
\be
\langle \gamma e\rangle _\theta = \langle \gamma_t({\bf
\theta_a}) e_t({\bf \theta_b})\rangle  + \langle \gamma_r({\bf
\theta_a}) e_r({\bf \theta_b})\rangle \,\,.
\ee
Note that we treat each realization of the N-body simulations as an
independent pointing and plot the mean and variance between the
results from each realisation.  This result represents the level of
contamination that one would expect from each galaxy model for a deep
lensing survey with a median redshift $z_m \sim 1$.  These results show
an anti-correlation, as predicted by \citet{HS04} and measured by
\citet{RM05}.  The strongest signal is found from the elliptical
model, where all occupied halos contain a galaxy whose shape is the
same as its parent dark matter halo.   The spiral model shows a
correlation that is consistent with zero on all scales, and the mix
model shows a weak but significant anti-correlation on scales
$\theta<20$ arcmin.  For reference, the solid theory curve shows the
the shear-shear correlation function (multiplied by $-1$) for a deep $z_s = 1$
survey using the adopted $\Lambda$CDM cosmology.  Note that 
the shear-ellipticity correlation for the mix model has a similar magnitude 
to the models tested in \citet{King05}.  In contrast however, we find signal 
that has power out to angular scales that are 
significantly larger than the $1$ arcmin cut-off considered in \citet{King05}.

Table~\ref{tab:contam} details the level of contamination to a weak lensing
measurement by a shear-ellipticity
correlation from each galaxy model, for both a deep $z_s \sim 1$ survey, whose
results are plotted in figure~\ref{fig:shearecor}, and medium deep
$z_s \sim 0.5$ survey.  The mix model shows a level of shear-ellipticity 
correlation which is $-6 \pm 1$\% of the lensing signal at an angular scale 
$\theta = 1$ arcmin for the deep $z_s \sim 1$ survey.

\begin{table}
\begin{center}
\begin{tabular}{l|c|c|c}
  $z_m \sim 1$ & $\theta \sim 0.5$ & $\theta \sim 1$ & $\theta \sim 10$ \\
  \hline
Elliptical & $  -19 \pm    1 $ \% & $  -20 \pm    1 $ \% & $  -31 \pm    2 $ \%  \\
Spirals    & $    1 \pm    1 $ \% & $    1 \pm    1 $ \% & $    2 \pm    2 $ \%  \\
Mix        & $   -6 \pm    1 $ \% & $   -6 \pm    1 $ \% & $  -10 \pm    2 $ \%  \\
 \hline
  $z_m \sim 0.5$ & $\theta \sim 0.5$ & $\theta \sim 1$ & $\theta \sim 10$ \\
  \hline
Elliptical & $  -67 \pm    5 $ \% & $  -43 \pm    5 $ \% & $  -82 \pm    8 $ \%  \\
Spirals    & $    8 \pm    5 $ \% & $    3 \pm    5 $ \% & $  -10 \pm    8 $ \%  \\
Mix        & $  -24 \pm    4 $ \% & $  -14 \pm    5 $ \% & $  -31 \pm    8 $ \%  \\

\end{tabular}
\end{center}
\caption{The percentage of contamination to a measurement 
of the shear-shear correlation function by a shear-ellipticity
correlation from each galaxy model at $\theta = 0.5, 1.0, 10.0$ arcmin.
Results are given for both a deep $z_s \sim 1$ lensing 
survey (upper rows), whose
results are plotted in figure~\ref{fig:shearecor}, and medium deep lensing
$z_s \sim 0.5$ survey (lower rows).} 
\label{tab:contam}
\end{table}

\begin{figure}
\begin{center}
\epsfig{file=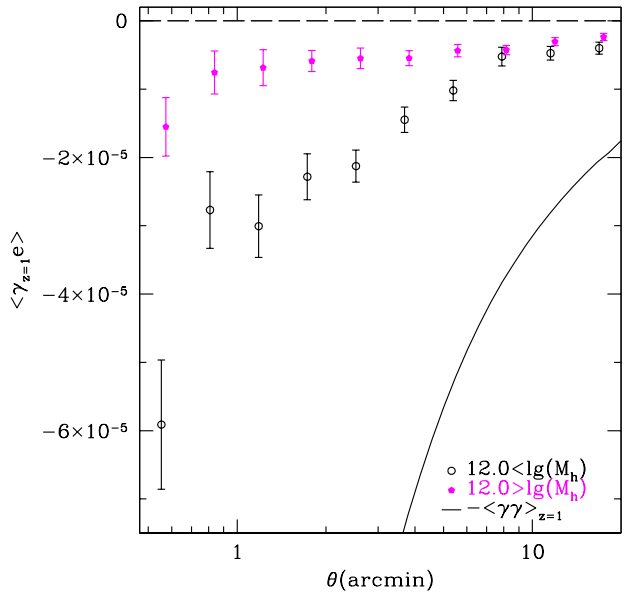,width=8.5cm,angle=0,clip=}
\caption{The shear-ellipticity correlation function $\langle \gamma e\rangle _\theta$ for the mix galaxy model for two samples of 
galaxies split by mass; where 
$M_h> 1.0\times 10^{12}h^{-1}{\rm M}_\odot$ (open) 
and where $M_h < 1.0\times 10^{12}h^{-1}{\rm M}_\odot$ (filled).  Note that 
all source galaxies (that experience a lensing 
shear $\gamma$) are placed at a source redshift $z_s = 1$ and that
for reference, the solid theory curve shows the
the shear-shear correlation function (multiplied by $-1$) for a deep $z_s = 1$
survey using the adopted $\Lambda$CDM cosmology.}
\label{fig:masscuts}
\end{center}
\end{figure}

Figure~\ref{fig:masscuts} shows the shear-ellipticity correlation function 
$\langle \gamma e \rangle$ for the mix galaxy model for two samples of 
galaxies split by mass.  
The strongest shear-ellipticity correlation is seen for the 
most massive galaxy sample where $M_h> 1.0\times 10^{12}h^{-1}{\rm M}_\odot$.  This  
is expected, as the strength of the lensing shear is proportional to the 
mass of the lens.  It is also reminiscent of what in seen in \citet{RM05}, 
where the strongest signal is measured from 
the most luminous and hence the most massive galaxy sample.  
This result allows us to briefly discuss 
the impact of multiply-occupied halos, 
which we have removed from our analysis as a result of our uncertainty in 
how to connect the properties of the parent dark matter halo to the 
luminous galaxy members.  One could consider the case where the central 
galaxy is represented by our elliptical model and the satellite 
galaxies are randomly oriented.  The multiply occupied halos are 
typically the more massive halos, and we would therefore expect to see a 
strong signal from the central galaxies, as seen with the most massive galaxy 
sample shown in figure~\ref{fig:masscuts}.  
Randomly oriented satellite galaxies would not contribute to the 
shear-ellipticity correlation signal and would thus dilute effect 
of the central galaxy.  The reality of this scenario is however 
difficult to quantify, and indeed there is evidence that suggests
satellite galaxies are not randomly oriented with a tendency 
to align with the major axis of the central galaxy \citep{Brainerd05,Yang06}.  
We therefore limit our discussion, 
in the rest of this paper to singly-occupied halos, deferring a 
quantitative analysis that includes satellite galaxies for a future paper 
analysing higher resolution N-body simulations.

\begin{figure}
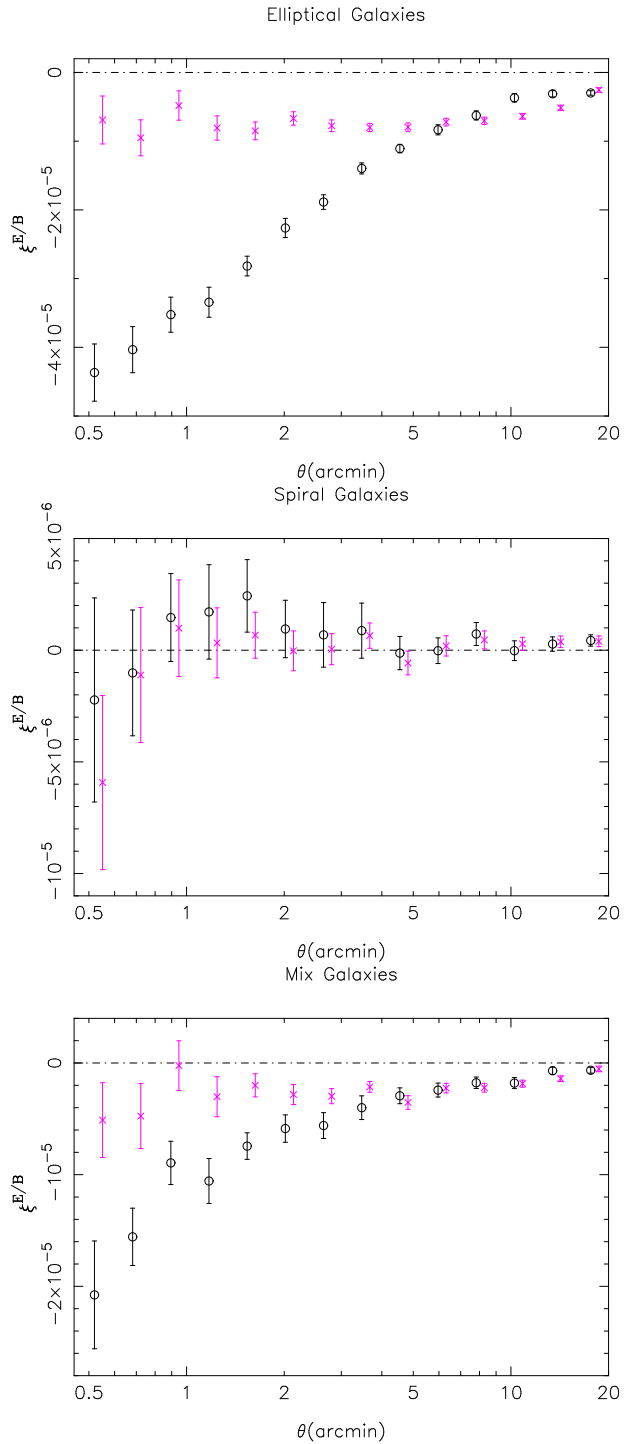

\begin{center}
\epsfig{file=EB_eli.ps,width=6.35cm,angle=270,clip=}
\epsfig{file=EB_sma.ps,width=6.35cm,angle=270,clip=}
\epsfig{file=EB_mix.ps,width=6.35cm,angle=270,clip=}
\end{center}
\caption{The E mode (circle) and B mode (cross) of the shear-ellipticity correlation function for the elliptical (upper panel), spiral (middle panel) and mix (lower panel) galaxy models.  For clarity the B mode points are slightly offset to the right.}
\label{fig:EB}
\end{figure}

Following \citet{CNPT02} and the method detailed in \citet{PenWM}, we
decompose the measured shear-ellipticity correlation functions into
their E and B modes, equation~\ref{eqn:xieb}, 
assuming that $\langle \gamma e\rangle _\theta = 0$ for
$\theta>20$ arcmin.  Figure~\ref{fig:EB} shows the resulting E and B
mode signals for all three galaxy models.   As expected the E and B
modes from the spiral model are consistent with zero.  For the mix and 
elliptical models, the B-modes are negative but weaker than the corresponding E-modes.  The B-modes are however fairly constant on scales 
$\theta<15$ arcmin and would therefore be detectable
if all B-modes from other
non-lensing sources, such as residual point-spread function
distortions, are negligible.

\subsection{Redshift dependence}
\label{sec:red_dep}

In this section we measure the dependence of the shear-ellipticity
correlation signal as a function of the foreground galaxy redshift.  
We present the results for the elliptical model as the strength of the 
signal allows us effectively to probe the redshift dependence in the
presence of increased noise from the redshift binning.  The trends seen
with the elliptical model are also seen with the mix model, 
although at lower significance, and the shear-ellipticity correlation 
for the spiral model remains consistent with zero for all redshift slices. 

\begin{figure}
\begin{center}
\epsfig{file=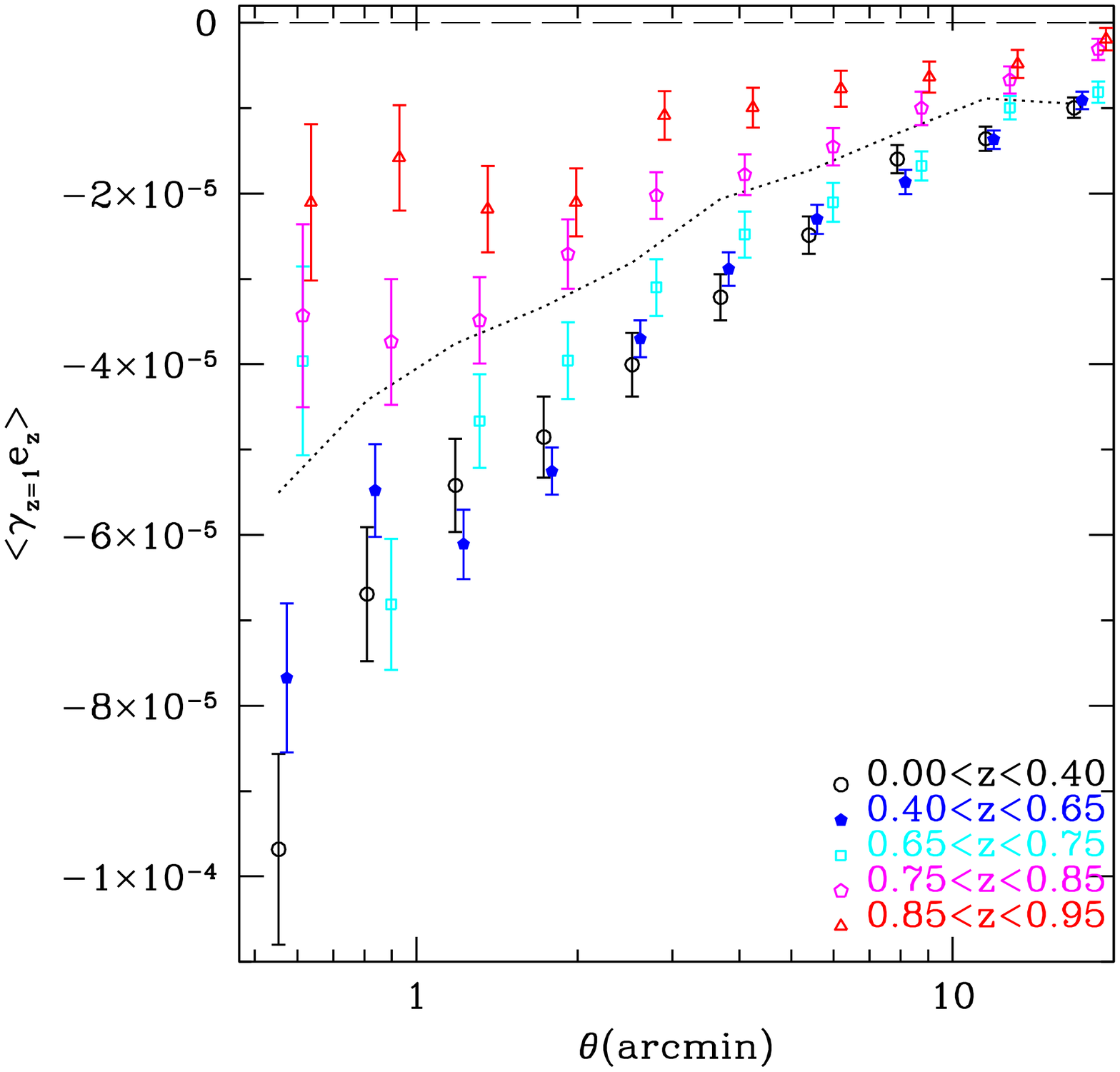,width=8.5cm,angle=0,clip=}
\epsfig{file=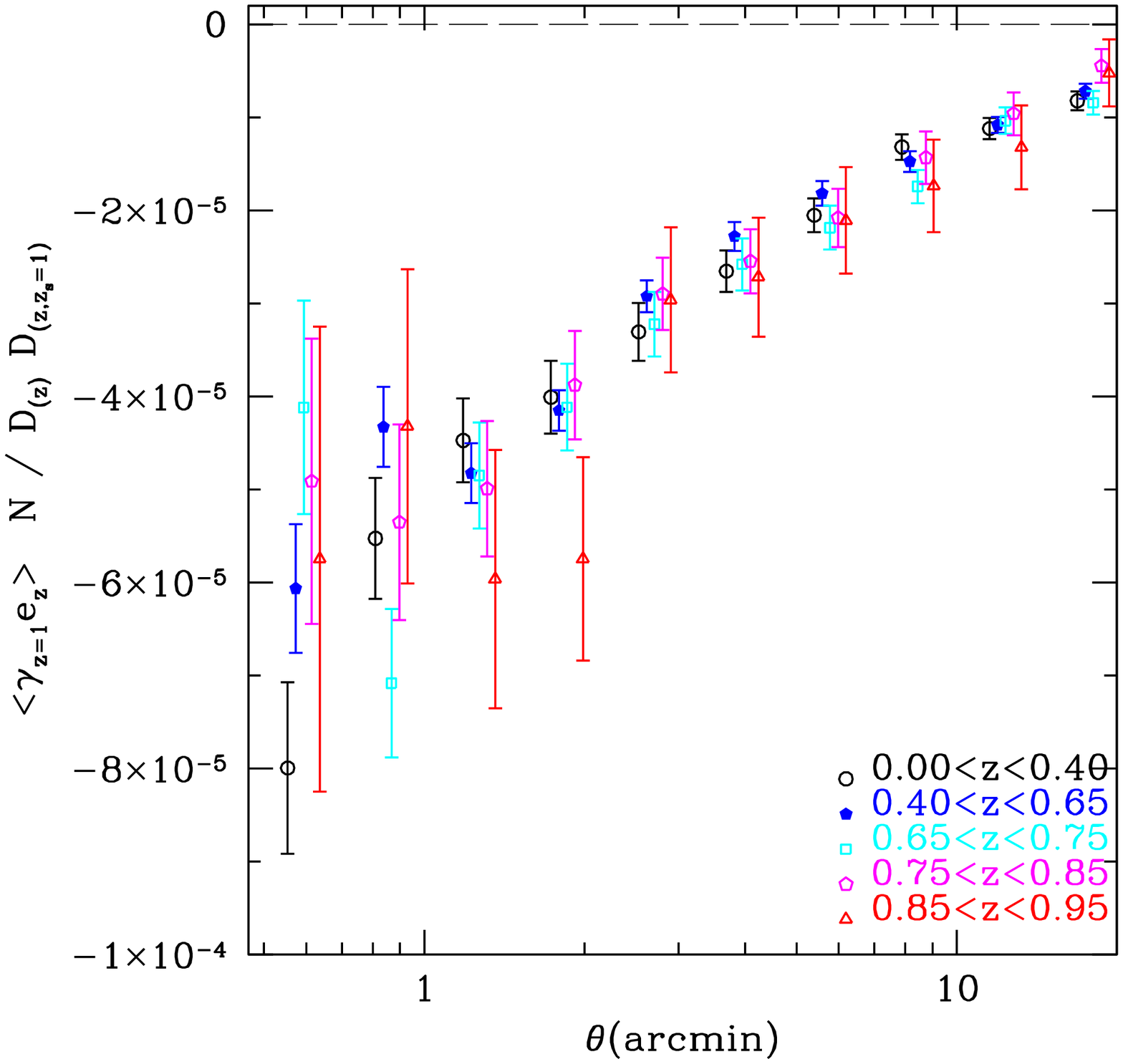,width=8.5cm,angle=0,clip=}
\end{center}
\caption{The redshift dependence of the shear-ellipticity correlation as a function of galaxy redshift slice $z_l$.  The upper panel shows the measured signal for the following galaxy redshifts; $0.0<z_l<0.4$ (circle),
$0.4<z_l<0.65$ (solid), $0.65<z_l<0.75$ (square), $0.75<z_l<0.85$ (hexagon), $0.85<z_l<0.95$ (triangle) .  The dashed line shows the elliptical measurement in Figure~\ref{fig:shearecor} from the full galaxy sample $0.0<z_l<1.0$.
The lower panel shows the same signal with the 
lensing efficiency $E$ 
factored out and normalized to the lensing efficiency $N$ of the
mean galaxy redshift of the full galaxy sample where $z_l=0.68$}
\label{fig:zevo}
\end{figure}

\begin{figure}
\begin{center}
\epsfig{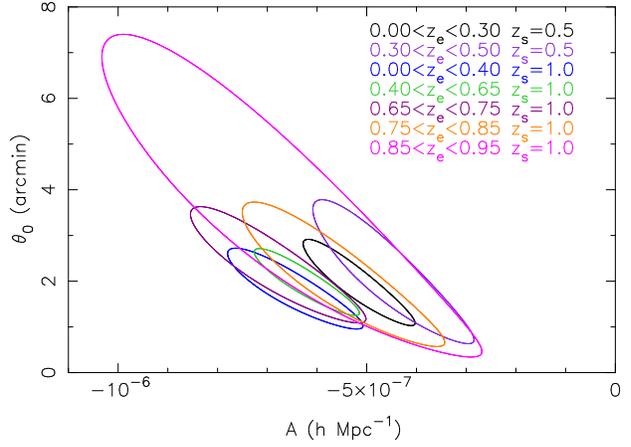}
\end{center}
\caption{$2\sigma$ constraints on the amplitude 
$A$ and scale dependence $\theta_0$ (equation~\ref{eqn:gefit}) 
of the shear-ellipticity correlation  assuming the elliptical galaxy model.
Parameter constraints are determined 
from the 5 different galaxy redshift slices shown in
figure~\ref{fig:zevo} where $\gamma$ is measured at $z_s = 1.0$, 
plus 2 additional 
low redshift slices where $\gamma$ is measured at $z_s = 0.5$.}
\label{fig:fits_evo}
\end{figure}

Figure~\ref{fig:zevo} shows the redshift dependent shear-ellipticity
correlation for 5 different galaxy redshift slices; $0.0-0.4$,
$0.4-0.65$, $0.65-0.75$, $0.75-0.85$, $0.85-0.95$.  
This result demonstrates  that the most
significant anti-correlation is found for the low redshift  ($z<0.65$)
galaxy samples, which are the most efficient for lensing source
galaxies at $z_s=1$.  The lensing efficiency $E$ of the
lens-source pair is defined to be
\be
E = \frac{D_l D_{ls}} {D_s} \,\, .
\label{eqn:lenseff}
\ee
where the distances $D$ are all comoving angular diameter
distances to the lens $D_l$, to the source $D_s$, and between the lens
and source $D_{ls}$.   
We might expect the shear-ellipticity correlation signal to scale with
the lensing efficiency and to test this hypothesis, the lower panel of
figure~\ref{fig:zevo} shows the shear-ellipticity correlation signal
with the lensing efficiency factored out; we divide by $E$ for the
median lens redshift in each redshift slice,  and multiply by a
normalising factor $N$ set equal to the lensing efficiency $E$ of the
mean redshift lens for the whole halo sample where 
$z_l=0.68$.  The result shows that the
redshift dependent 
shear-ellipticity correlation measurements are very
consistent with one another once the lensing efficiency is factored out.
This then
motivates the following parameterisation
of the redshift evolution of
the shear-ellipticity correlation,
\be
\langle \gamma(z_s) \, e(z_l)\rangle _\theta = E \, \frac{A}{\theta+\theta_0} \,\, ,
\label{eqn:gefit}
\ee
where the amplitude $A$ and scale dependence $\theta_0$ are free parameters.  
$E$ is the lensing efficiency of the lens-source ($\gamma e$) pair 
(equation~\ref{eqn:lenseff}), and $z_l$ is set equal to the median redshift of 
each galaxy redshift slice.  Note that this parameterisation is 
similar to the exponential
parameterisation proposed by \citet{King05}, but this functional
form is found to produce 
a better fit to the results.
Figure~\ref{fig:fits_evo} shows the two parameter $2\sigma$ 
constraints on the amplitude 
$A$ and scale dependence $\theta_0$ as measured independently 
from the 5 different 
galaxy redshift slices shown in
figure~\ref{fig:zevo} where $\gamma$ is measured at $z_s = 1.0$, 
plus 2 additional 
low redshift slices where $\gamma$ is measured at $z_s = 0.5$.
The consistent results for each galaxy redshift slice 
suggest that the evolution of the shear-ellipticity 
correlation can be safely modelled through the lensing efficiency of the 
lens-source pair.

\begin{figure}
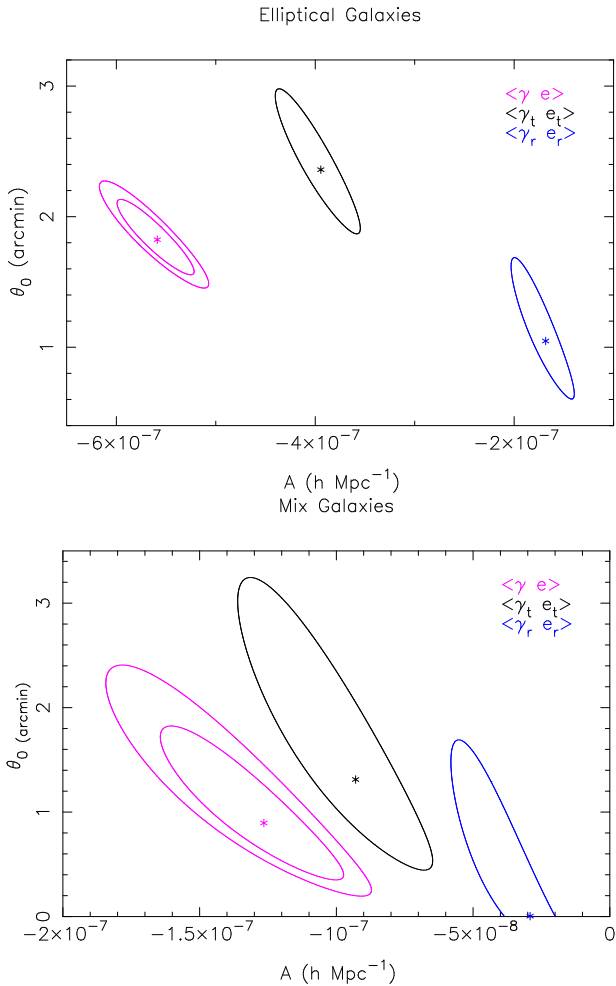

\begin{center}
\epsfig{file=GI_eli_fits.ps,width=6.5cm,angle=270,clip=}
\epsfig{file=GI_mix_fits.ps,width=6.5cm,angle=270,clip=}
\end{center}
\caption{$1$ and $2\sigma$ constraints on the amplitude 
$A$ and scale dependence $\theta_0$ (equation~\ref{eqn:gefit}) 
of the shear-ellipticity correlation $\langle \gamma e \rangle$ , 
assuming the elliptical galaxy model (upper) and mix galaxy model (lower). 
For clarity, only the $1\sigma$ constraints are shown 
for the tangential $\langle \gamma_t e_t\rangle$ and 
rotated $\langle \gamma_r e_r\rangle$ shear-ellipticity correlations, 
where $\langle \gamma_r e_r\rangle$ is the weakest signal. 
The best-fit parameters are marked with a star.}
\label{fig:fits}
\end{figure}

Using this parameterisation (equation~\ref{eqn:gefit}) we place
constraints on the amplitude $A$ and scale dependence $\theta_0$ of the 
shear-ellipticity correlation measured jointly from the redshift dependent
results. Figure~\ref{fig:fits} shows the resulting two-parameter $1$ and 
$2\sigma$
constraints for the elliptical model (upper) and mix model (lower).
The best fitting results for the elliptical model are 
$A = (-5.60 \pm 0.18) \times 10^{-7} h \, {\rm Mpc}^{-1}$ 
and $\theta_0 = 1.83 \pm 0.13$ arcmin.  
For the mix model 
$A = (-1.29 \pm 0.16) \times 10^{-7} h \, {\rm Mpc}^{-1}$ and 
$\theta_0 = 0.93^{+ 0.38}_{-0.29} $ arcmin.  The $1\sigma$
errors on $A$ are marginalised over $\theta_0$ and vice versa. 
The constraints measured for the tangential $\langle \gamma_t e_t\rangle$ and 
rotated $\langle \gamma_r e_r\rangle$ shear-ellipticity correlations are 
also shown in figure~\ref{fig:fits}, 
but for clarity only the $1\sigma$ constraints are presented.  Note that the 
$\langle \gamma_r e_r\rangle$ shear-ellipticity correlation is always weaker 
than the $\langle \gamma_t e_t\rangle$ shear-ellipticity correlation.

\section{Discussion}
\label{sec:discussion}

The results that have been presented in this paper and others 
(e.g H04, \citet{HS04,RM05}), demonstrate
that if a relation exists between the shape of galaxies and the
properties of their parent dark matter halos then there is the
potential for contamination to the measurement of weak gravitational
lensing by large-scale structure.  The conclusion that one should draw
from these works is that it is no longer acceptable to assume that all
the intrinsic contributions to the observed galaxy ellipticity 
correlation function (i.e. terms that
include $e_{\rm s}$ in
equation~\ref{eqn:ellicor}) are negligible.  We must therefore seek to derive
and apply optimal methods to remove these sources of contamination in 
the analyses of the next generation of weak lensing surveys.
In this paper we have sought to answer the question, what level of
contamination might one expect, and our conclusions are clearly
dependent on our choice of galaxy model; ellipticals, spirals or mix.

A wealth of observational evidence allows us immediately to rule out the
elliptical model on several grounds.  The measurement of the intrinsic
alignment signal $\eta(r)$ from the SDSS \citep{RM05} is inconsistent
with the elliptical model signal at the 99.9\% level, as shown in
Figure~\ref{fig:IA_nr}.  Similar constraints could also be placed from the
intrinsic alignment signal measured in the SuperCOSMOS survey \citep{MLB02}.
Figure~\ref{fig:IA_theta} shows the expected
angular ellipticity correlation in the absence of weak lensing for a
survey with median redshift $z_m = 0.7$.  This medium depth survey 
is similar to the Red
sequence Cluster Survey (RCS) \citet{HYG02} and the CTIO survey
\citep{Jarvis}.  Weak lensing analyses of these two surveys constrain
the amplitude of the matter  power spectrum to be $\sigma_8 \sim 0.8
\pm 0.15 $ for $\Omega_m = 0.3$,  which is consistent with the results
from the analysis of the cosmic microwave background \citep{WMAP06}.
If more than half of the RCS and CTIO 
measured lensing signal came from intrinsic
alignments, as suggested by the results of the elliptical model, 
the `true' value of $\sigma_8$ would be then be very low and
inconsistent with other observational constraints.  The COMBO-17 survey has 
been analysed using redshift weighting to remove the intrinsic contribution 
from close galaxy pairs (H04) and only a small difference was noted in the 
resulting signal, placing constraints on the intrinsic alignment contamination 
that would also reject the elliptical model at the 99\% confidence level.
Lastly, if the shear-ellipticity correlations were as
strongly anti-correlated as expected from an elliptical model for
median-deep surveys (see Table~\ref{tab:contam}) 
we would expect to measure no weak lensing signal in the RCS and CTIO 
surveys, which is not the case.  All of the evidence therefore 
suggests that not every galaxy has the same shape as its parent halo, 
and we can therefore rule out the 100\% elliptical model.

The results from the spiral model are consistent with the \citet{RM05} 
SDSS constraints on intrinsic galaxy alignment. 
It predicts a very low intrinsic alignment correlation 
and low shear-ellipticity correlation that is consistent with zero on all 
scales. It is important however to take note of  
the SDSS results for their `L6' massive elliptical galaxy sample 
\citep{RM05}. This sample of massive ellipticals
exhibit a strong correlation between their galaxy shape and the surrounding 
density field.  This observation has been propagated to high redshift to 
predict a measured
shear-ellipticity anti-correlation that is $\sim 100\% \pm 50\%$ of
the magnitude 
of the expected shear-shear correlation signal for a survey with $z_m = 1$.  
The amplitude of the `L6' shear-ellipticity correlation is 
similar to the measured shear-ellipticity correlation 
signal from our elliptical model, but is 
inconsistent at the $4\sigma$ level with
the measured shear-ellipticity correlation 
signal from our spiral model.
In order to reproduce the strong SDSS signal from a sub-sample 
of galaxies in our simulations we therefore require
a small population of elliptical galaxies and can thus rule out 
the 100\% spiral model. 

We have modelled a mixed population of spiral and elliptical galaxies using
the early-type/late-type halo occupation distribution 
of \citet{Cooray06} which reproduces the observed 
luminosity functions and clustering properties of red and blue galaxies.
The results from this mix model are consistent with the all 
the observational results of \citet{RM05}
It predicts low intrinsic alignment correlation,
weak, though significant, shear-ellipticity correlations, and negative
B-modes.  B-modes from systematics associated with the observations
(for example residual distortions from the optical point-spread
function) would typically produce positive B-modes and hence one would
not expect to be able to detect the weakly negative B-mode signature
of the mix model in current data.  

The effect of the shear-ellipticity correlation is to reduce the 
measured shear-shear correlation below the amplitude of the 
true cosmological lensing signal.  
If uncorrected for, this would 
result in an underestimate of $\sigma_8$ from weak lensing measurements.  
The degree of underestimation would be survey depth dependent as
from Table~\ref{tab:contam} we expect a medium-depth survey to
suffer $\sim 20\%$ contamination and a deep survey to suffer $\sim 10\%$
contamination. A comparison of the results from
different depth surveys should therefore, in principle, be able to constrain 
models of shear-ellipticity correlations as one would expect to measure lower
values of $\sigma_8$ from shallower surveys. 
Whilst it is tempting to discuss the trend in the literature
of somewhat higher $\sigma_8$ measurements from deeper lensing surveys, the 
uncertainties in the source redshift distribution \citep{VW06}, 
sampling variance
and shear calibration errors \citep{HymzSTEP} 
means that we are unable to make any firm
conclusions with current lensing data to confirm or rule out the mix galaxy 
model.

\subsection{Diagnostics for shear-ellipticity contamination}

For the next generation of lensing surveys, the technical systematics
associated with the detection and measurement of weak lensing have a
promising chance of being negligible.  A significant detection of
negative B-modes would then suggest contamination from
shear-ellipticity correlations.  In addition,  the results of
\citet{RM05} and the comparison of the different models used  in this
analysis demonstrate that shear-ellipticity correlation is likely to
be galaxy morphology dependent.  For future  multi-colour surveys with
photometric  redshifts one will be able effectively to separate red-sequence
elliptical galaxies  from the galaxy sample using a simple colour cut
(see for example \citet{Bell04}).  With large area surveys one could
then  analyse the data with and without the
elliptical galaxy population.  A difference between the two measurements
would suggest  contamination from shear-ellipticity correlations.  If
that is the case,  a zero B-mode measurement from an analysis where all
elliptical-type galaxies have been rejected 
might well serve as a good indication that the
contamination has been removed.

In section~\ref{sec:red_dep} we investigated the redshift dependence
of the shear-ellipticity correlation 
$\langle \gamma_\alpha e_\beta\rangle $ where
$z_\alpha < z_\beta$.  The results, shown in figure~\ref{fig:zevo}
demonstrated that the evolution of the signal is proportional to the
lensing efficiency $E$ (equation~\ref{eqn:lenseff}) 
of the lens-source pair, where, for
the shear-ellipticity correlation $\langle \gamma_\alpha
e_\beta\rangle $ the source is at $z_\alpha$ and the lens is at
$z_\beta$. 
The model that
provides the most realistic description appears to be the mix model for which
we have constrained the amplitude of the $\langle \gamma e \rangle $ signal
to $\sim 10\%$ accuracy.  
We have calculated the level of contamination arising from a mixed 
galaxy population for a deep $z_m \sim 1$
survey analysed using the shear correlation function 
$\langle \gamma \gamma \rangle$, and found it to be at the level of 
$-6 \pm 1$\% and $-10 \pm  2$\% at angular scales 1 and 10 arcmin 
respectively.  
Using the $\langle \gamma e \rangle $ model (equation~\ref{eqn:gefit}) 
with the best fitting parameters to our results,
one could now potentially remove this contaminating signal 
to an accuracy of better than 
$1$ \% of the lensing signal.
We would urge caution with applying such a method however, as our results in 
section~\ref{sec:voids} 
clearly indicate the limitations of the simple models 
that we have used to link the properties of galaxies with their parent 
dark matter halos.  We would therefore recommend using the 
$\langle \gamma e \rangle$ 
model (equation~\ref{eqn:gefit}) 
that we have provided as a guide to the level of contamination, 
which should be self-consistently determined from the lensing data using 
the result that the contamination scales with lensing efficiency.
As proposed by \citet{HS04}, measurements of cross-correlation
tomography $ \langle \gamma_\alpha \gamma_\beta\rangle $ 
(see for example \citet{TakadaWhite}) will be a
useful diagnostic tool, as the true cosmological lensing signal scales
differently with redshift with the
lensing efficiency integrated over all foreground lenses.
One would 
therefore able to project out the different redshift dependences of the 
two signals as demonstrated by \citet{King05} and 
simultaneously constrain the cosmological lensing signal and 
the contaminating shear ellipticity correlations.

\section{Conclusion}
\label{sec:conc}

We have investigated potential sources of contamination to the
measurement of weak lensing by large-scale structure from the
intrinsic alignment of galaxies, in section~\ref{sec:IAres}, and from
the presence of shear-ellipticity correlations between foreground
galaxy shapes and the weak lensing shear experienced by background
galaxies, in section~\ref{sec:res}.  These two sources of physical
contamination behave rather differently.  The intrinsic alignment
contamination arises from close galaxy pairs, and can thus be
eliminated by downweighting those galaxy pairs which are at close
separations \citep{HH03,KingSch02}, or by measuring cross-correlation
tomography between different redshift bins.  
Intrinsic ellipticity-ellipticity correlations are thus not a problem 
if photometric redshifts are available. 
In contrast, 
the shear-ellipticity correlation arises from galaxy pairs which are at
large physical separations, with the strongest correlations arising
for lens-source pairs where the lensing efficiency $E$ 
(equation~\ref{eqn:lenseff}) is at its
strongest.  This could potentially be directly 
eliminated through morphological
galaxy selection, or from measuring weak lensing shear
correlation within thin redshift slices \citep{King05}.  All possible 
techniques to remove these sources of contamination will require good 
photometric redshift measurements and acquiring deep multi-colour data 
should be made a top priority for all future lensing surveys.

As discussed in section~\ref{sec:discussion} the conclusions are
dependent on the choice of model that we use to connect galaxy shape
to the properties of its parent dark matter halo.  General trends are
that for deeper surveys, where the cosmological lensing signal is at
its strongest, the degree of contamination is decreased.  Eliminating
the elliptical model through comparison to several observation results
allows us conclude that for medium deep surveys $z_m>0.5$, intrinsic
alignment contamination contributes only to the small angular scales
$\theta<1.5$ arcmin, and we refer the reader to H04 for a more
detailed discussion on the implications of intrinsic galaxy alignments
for future lensing surveys.  The shear-ellipticity  correlation
is consistent with zero for our spiral model.  To reproduce the
results of \citet{RM05} however it is necessary to include a
population of elliptical galaxies which introduce contamination to the
shear lensing signal at the $\sim 10\%$ level for a deep $z_m = 1$
survey and $\sim 20\%$ level for a medium deep $z_m = 0.5$ survey (see
table~\ref{tab:contam}).

The galaxy models that we have applied to the simulations are
indeed clear simplifications of reality as they consider only the properties
of the parent halo, not the halo environment or the 
tidal density field directly.  
The non-detection of any alignment between a galaxy's spin axis and
the surface of voids in our simulations, in section~\ref{sec:voids},
is in some tension with the analysis of the 2dF and SDSS
surveys \citep{Trujillo}.  If the observation is robust, it is therefore 
likely that it is not only the properties
of the parent halo that impact on galaxy shape, but also the
surrounding environment to some extent.  In the rare cases of galaxies near
voids or clusters, we might well expect the environment to have a 
stronger effect on galaxy shape than its parent dark matter halo and it 
will be important to investigate this further
with full hydro-dynamical simulations on cosmological scales 
\citep[see for example][]{NAB04}.  

The important point to take away from this analysis is that our
numerical study supports the notion that the contamination from 
shear-ellipticity correlation scales with lensing efficiency. 
This simple scaling with redshift offers possibilities of 
determining the cosmic shear signal and the contamination 
simultaneously, either by using the models constrained in 
section~\ref{sec:red_dep} for the behaviour of the contamination 
\citep[cf][]{King05}, or by making fewer assumptions and parameterising 
the contamination.  Because of the redshift scaling, 
only a single function needs 
to be parameterised. Cosmological parameter estimation would then proceed by 
marginalising over the nuisance parameters describing the contamination.
Within the context of the model considered here, the $\langle \gamma e \rangle$
contamination can be estimated to an accuracy of 3\% and 10\% for the 
elliptical and mix models respectively.  

A more detailed study of the redshift dependence of 
shear-ellipticity correlation will, however, be required for the more 
ambitious future lensing experiments which aim to determine $w(z)$. 
Such studies will need to include the satellite galaxy population that is 
missing from this analysis in addition to an 
improved mixed galaxy population model that takes into account 
an observationally constrained 
redshift evolution in the ratio of elliptical to spiral galaxies 
for different halo masses.  This will require the development of better 
models for galaxy formation in dark matter simulations to
improve our understanding of the connection between galaxies and 
their dark matter halos and the resulting impact on 
weak lensing studies.  In the meanwhile we should rely on probing this 
phenomenon with existing data sets, in order to develop suitable diagnostics 
and analysis tools that will not subject weak lensing measurements
to the kinds of contamination investigated in this paper thus allowing weak 
lensing to reach its full potential as a high precision dark matter and 
dark energy probe.

\section{Acknowledgments}
We thank Rachel Mandelbaum and Chris Hirata for useful discussions on 
this project, Asantha Cooray for making code based on his CLF 
publicly available, and Ignacio Trujillo for providing us with 
his results and the analytical predictions of \citet{Lee04}.
We would also like to thank the referee for his/her helpful comments.
The simulations used in this analysis 
were performed on the 
IBM-SP at NERSC.  CH is supported by a CITA National fellowship
and, along with LVW, acknowledges support from NSERC and CIAR.  
MW was supported in part by NASA and the NSF.

\bibliographystyle{mn2e}
\bibliography{/Users/Hymz/GEMS/galgal/paper/ceh_2005}

\begin{thebibliography}{}

\bibitem[\protect\citeauthoryear{Bacon, Massey, Refregier \& Ellis}{Bacon
  et~al.}{2003}]{BMRE}
Bacon D.,  Massey R.,  Refregier A.,    Ellis R.,  2003, MNRAS, 344, 673

\bibitem[\protect\citeauthoryear{{Bell}, {Wolf}, {Meisenheimer}, {Rix},
  {Borch}, {Dye}, {Kleinheinrich}, {Wisotzki} \& {McIntosh}}{{Bell}
  et~al.}{2004}]{Bell04}
{Bell} E.~F.,  {Wolf} C.,  {Meisenheimer} K.,  {Rix} H.,  {Borch} A.,  {Dye}
  S.,  {Kleinheinrich} M.,  {Wisotzki} L.,    {McIntosh} D.~H.,  2004, ApJ,
  608, 752

\bibitem[\protect\citeauthoryear{Bernstein \& Jarvis}{Bernstein \&
  Jarvis}{2002}]{Bernstein}
Bernstein G.~M.,  Jarvis M.,  2002, AJ, 123, 583

\bibitem[\protect\citeauthoryear{{Brainerd}}{{Brainerd}}{2005}]{Brainerd05}
{Brainerd} T.~G.,  2005, ApJL, 628, L101

\bibitem[\protect\citeauthoryear{Brown, Taylor, Bacon, Gray, Dye, Meisenheimer
  \& Wolf}{Brown et~al.}{2003}]{MLB02}
Brown M.,  Taylor A.,  Bacon D.,  Gray M.,  Dye S.,  Meisenheimer K.,    Wolf
  C.,  2003, MNRAS, 341, 100

\bibitem[\protect\citeauthoryear{Brown, Taylor, Hambly \& Dye}{Brown
  et~al.}{2002}]{BTHD02}
Brown M.,  Taylor A.,  Hambly N.,    Dye S.,  2002, MNRAS, 333, 501

\bibitem[\protect\citeauthoryear{{Catelan}, {Kamionkowski} \&
  {Blandford}}{{Catelan} et~al.}{2001}]{CKB01}
{Catelan} P.,  {Kamionkowski} M.,    {Blandford} R.~D.,  2001, MNRAS, 320, L7

\bibitem[\protect\citeauthoryear{{Cooray}}{{Cooray}}{2006}]{Cooray06}
{Cooray} A.,  2006, MNRAS, 365, 842

\bibitem[\protect\citeauthoryear{{Cooray} \& {Milosavljevi{\'c}}}{{Cooray} \&
  {Milosavljevi{\'c}}}{2005}]{CLF_CM05}
{Cooray} A.,  {Milosavljevi{\'c}} M.,  2005, ApJL, 627, L89

\bibitem[\protect\citeauthoryear{Crittenden, Natarajan, Pen \&
  Theuns}{Crittenden et~al.}{2001}]{CNPT01}
Crittenden R.,  Natarajan R.,  Pen U.,    Theuns T.,  2001, ApJ, 559, 552

\bibitem[\protect\citeauthoryear{Crittenden, Natarajan, Pen \&
  Theuns}{Crittenden et~al.}{2002}]{CNPT02}
Crittenden R.,  Natarajan R.,  Pen U.,    Theuns T.,  2002, ApJ, 568, 20

\bibitem[\protect\citeauthoryear{{Croft} \& {Metzler}}{{Croft} \&
  {Metzler}}{2000}]{CM00}
{Croft} R.~A.~C.,  {Metzler} C.~A.,  2000, ApJ, 545, 561

\bibitem[\protect\citeauthoryear{{Davis}, {Efstathiou}, {Frenk} \&
  {White}}{{Davis} et~al.}{1985}]{DEFW}
{Davis} M.,  {Efstathiou} G.,  {Frenk} C.~S.,    {White} S.~D.~M.,  1985, ApJ,
  292, 371

\bibitem[\protect\citeauthoryear{{Hamana}, {Miyazaki}, {Shimasaku}, {Furusawa},
  {Doi}, {Hamabe}, {Imi}, {Kimura}, {Komiyama}, {Nakata}, {Okada}, {Okamura},
  {Ouchi}, {Sekiguchi}, {Yagi} \& {Yasuda}}{{Hamana} et~al.}{2003}]{Hamana}
{Hamana} T.,  {Miyazaki} S.,  {Shimasaku} K.,  {Furusawa} H.,  {Doi} M.,
  {Hamabe} M.,  {Imi} K.,  {Kimura} M.,  {Komiyama} Y.,  {Nakata} F.,  {Okada}
  N.,  {Okamura} S.,  {Ouchi} M.,  {Sekiguchi} M.,  {Yagi} M.,    {Yasuda} N.,
  2003, ApJ, 597, 98

\bibitem[\protect\citeauthoryear{{Heavens} \& {Peacock}}{{Heavens} \&
  {Peacock}}{1988}]{HP88}
{Heavens} A.,  {Peacock} J.,  1988, MNRAS, 232, 339

\bibitem[\protect\citeauthoryear{{Heavens}, {Refregier} \& {Heymans}}{{Heavens}
  et~al.}{2000}]{HRH00}
{Heavens} A.,  {Refregier} A.,    {Heymans} C.,  2000, MNRAS, 319, 649

\bibitem[\protect\citeauthoryear{Heymans, {Bell}, {Rix}, {Barden}, {Borch},
  {Caldwell}, {McIntosh}, {Meisenheimer}, {Peng}, {Wolf}, {Beckwith}, ,
  {H\"au\ss ler}, {Jahnke}, {Jogee}, , {Sanchez}, {Somerville} \&
  {Wisotzki}}{Heymans et~al.}{2006}]{GEMSgg}
Heymans C.,  {Bell} E.~F.,  {Rix} H.-W.,  {Barden} M.,  {Borch} A.,  {Caldwell}
  J.~A.~R.,  {McIntosh} D.~H.,  {Meisenheimer} K.,  {Peng} C.~Y.,  {Wolf} C.,
  {Beckwith} S.~V.~W.,   {H\"au\ss ler} B.,  {Jahnke} K.,  {Jogee} S.,
  {Sanchez} S.~F.,  {Somerville} R.~S.,    {Wisotzki} L.,  2006, MNRAS Letters
  accepted

\bibitem[\protect\citeauthoryear{Heymans, Brown, Heavens, Meisenheimer, Taylor
  \& Wolf}{Heymans et~al.}{2004}]{HBH04}
Heymans C.,  Brown M.,  Heavens A.,  Meisenheimer K.,  Taylor A.,    Wolf C.,
  2004, MNRAS, 347, 895

\bibitem[\protect\citeauthoryear{Heymans, Brown, {Barden}, {Caldwell},
  {Jahnke}, {Rix}, Taylor, {Beckwith}, {Bell}, {Borch}, {H\"au\ss ler},
  {Jogee}, {McIntosh}, {Meisenheimer}, {Peng}, {Sanchez}, {Somerville},
  {Wisotzki} \& {Wolf}}{Heymans et~al.}{2005}]{HymzGEMS}
Heymans C.,  Brown M.~L.,  {Barden} M.,  {Caldwell} J.~A.~R.,  {Jahnke} K.,
  {Rix} H.-W.,  Taylor A.~N.,  {Beckwith} S.~V.~W.,  {Bell} E.~F.,  {Borch} A.,
   {H\"au\ss ler} B.,  {Jogee} S.,  {McIntosh} D.~H.,  {Meisenheimer} K.,
  {Peng} C.~Y.,  {Sanchez} S.~F.,  {Somerville} R.~S.,  {Wisotzki} L.,
  {Wolf} C.,  2005, MNRAS, 160

\bibitem[\protect\citeauthoryear{Heymans \& Heavens}{Heymans \&
  Heavens}{2003}]{HH03}
Heymans C.,  Heavens A.,  2003, MNRAS, 339, 711

\bibitem[\protect\citeauthoryear{{Heymans et al.}}{{Heymans et
  al.}}{2006}]{HymzSTEP}
{Heymans et al.} C.,  2006, MNRAS, 368, 1323

\bibitem[\protect\citeauthoryear{{Hirata} \& {Seljak}}{{Hirata} \&
  {Seljak}}{2004}]{HS04}
{Hirata} C.~M.,  {Seljak} U.,  2004, Phys. Rev. D, 70

\bibitem[\protect\citeauthoryear{Hoekstra, Mellier, Waerbeke, Fu, Hudson,
  Maoli, Parker, Tereno \& Benabed}{Hoekstra et~al.}{2006}]{CFHTLS_wide}
Hoekstra H.,  Mellier Y.,  Waerbeke L.~V.,  Fu E. S.~L.,  Hudson M.~J.,  Maoli
  R.,  Parker L.,  Tereno I.,    Benabed K.,  2006, A\&A, accepted,
  astroph/0511089

\bibitem[\protect\citeauthoryear{Hoekstra, Yee \& Gladders}{Hoekstra
  et~al.}{2002}]{HYG02}
Hoekstra H.,  Yee H.,    Gladders M.,  2002, ApJ, 577, 595

\bibitem[\protect\citeauthoryear{Hoyle}{Hoyle}{1949}]{Hoyle}
Hoyle F.,  1949, Problems of Cosmical Aerodynamics, p.~195

\bibitem[\protect\citeauthoryear{Hu}{Hu}{1999}]{Hu99}
Hu W.,  1999, ApJ, 552, L21

\bibitem[\protect\citeauthoryear{Jarvis, Bernstein, Jain, Fischer, Smith, Tyson
  \& Wittman}{Jarvis et~al.}{2003}]{Jarvis}
Jarvis M.,  Bernstein G.,  Jain B.,  Fischer P.,  Smith D.,  Tyson J.,
  Wittman D.,  2003, ApJ, 125, 1014

\bibitem[\protect\citeauthoryear{{Jenkins}, {Frenk}, {Thomas}, {Colberg},
  {White}, {Couchman}, {Peacock}, {Efstathiou} \& {Nelson}}{{Jenkins}
  et~al.}{1998}]{Jenkins}
{Jenkins} A.,  {Frenk} C.,  {Thomas} P.,  {Colberg} J.,  {White} S.,
  {Couchman} H.,  {Peacock} J.,  {Efstathiou} G.,    {Nelson} A.,  1998, ApJ,
  499, 20

\bibitem[\protect\citeauthoryear{Jing}{Jing}{2002}]{Jing}
Jing Y.~P.,  2002, MNRAS, 335, L89

\bibitem[\protect\citeauthoryear{{Kaiser} \& {Squires}}{{Kaiser} \&
  {Squires}}{1993}]{KS93}
{Kaiser} N.,  {Squires} G.,  1993, ApJ, 404, 441

\bibitem[\protect\citeauthoryear{King \& Schneider}{King \&
  Schneider}{2002}]{KingSch02}
King L.,  Schneider P.,  2002, A\&A, 396, 411

\bibitem[\protect\citeauthoryear{{King}}{{King}}{2005}]{King05}
{King} L.~J.,  2005, A\&A, 441, 47

\bibitem[\protect\citeauthoryear{Lee}{Lee}{2004}]{Lee04}
Lee J.,  2004, ApJ, 614, L1

\bibitem[\protect\citeauthoryear{{Lee} \& {Pen}}{{Lee} \& {Pen}}{2001}]{LP01}
{Lee} J.,  {Pen} U.,  2001, ApJ, 555, 106

\bibitem[\protect\citeauthoryear{{Lee}, {Pen} \& {Seljak}}{{Lee}
  et~al.}{2002}]{LP02}
{Lee} J.,  {Pen} U.,    {Seljak} U.,  2002, ApJ, 543, L107

\bibitem[\protect\citeauthoryear{Mackey, White \& Kamionkowski}{Mackey
  et~al.}{2002}]{Mackey02}
Mackey J.,  White M.,    Kamionkowski M.,  2002, MNRAS, 332, 788

\bibitem[\protect\citeauthoryear{{Mandelbaum}, {Hirata}, {Ishak}, {Seljak} \&
  {Brinkmann}}{{Mandelbaum} et~al.}{2006}]{RM05}
{Mandelbaum} R.,  {Hirata} C.~M.,  {Ishak} M.,  {Seljak} U.,    {Brinkmann} J.,
   2006, MNRAS, 367, 611

\bibitem[\protect\citeauthoryear{{Massey}, {Refregier}, {Bacon}, {Ellis} \&
  {Brown}}{{Massey} et~al.}{2005}]{Massey}
{Massey} R.,  {Refregier} A.,  {Bacon} D.~J.,  {Ellis} R.,    {Brown} M.~L.,
  2005, MNRAS, 359, 1277

\bibitem[\protect\citeauthoryear{{Navarro}, {Abadi} \& {Steinmetz}}{{Navarro}
  et~al.}{2004}]{NAB04}
{Navarro} J.~F.,  {Abadi} M.~G.,    {Steinmetz} M.,  2004, ApJL, 613, L41

\bibitem[\protect\citeauthoryear{{Patiri}, {Betancort-Rijo}, {Prada}, {Klypin}
  \& {Gottl{\"o}ber}}{{Patiri} et~al.}{2006}]{Patiri}
{Patiri} S.~G.,  {Betancort-Rijo} J.~E.,  {Prada} F.,  {Klypin} A.,
  {Gottl{\"o}ber} S.,  2006, MNRAS, 369, 335

\bibitem[\protect\citeauthoryear{{Peebles}}{{Peebles}}{1969}]{P69}
{Peebles} J.,  1969, ApJ, 155, 393

\bibitem[\protect\citeauthoryear{Pen, {Van Waerbeke} \& Mellier}{Pen
  et~al.}{2002}]{PenWM}
Pen U.-L.,  {Van Waerbeke} L.,    Mellier Y.,  2002, ApJ, 567, 31

\bibitem[\protect\citeauthoryear{Porciani, Dekel \& Hoffman}{Porciani
  et~al.}{2002}]{Porciani}
Porciani C.,  Dekel A.,    Hoffman Y.,  2002, MNRAS, 332, 325

\bibitem[\protect\citeauthoryear{{Rhodes}, {Refregier}, {Collins}, {Gardner},
  {Groth} \& {Hill}}{{Rhodes} et~al.}{2004}]{RhodesSTIS}
{Rhodes} J.,  {Refregier} A.,  {Collins} N.~R.,  {Gardner} J.~P.,  {Groth}
  E.~J.,    {Hill} R.~S.,  2004, ApJ, 605, 29

\bibitem[\protect\citeauthoryear{{Rhodes}, {Refregier} \& {Groth}}{{Rhodes}
  et~al.}{2000}]{RRG00}
{Rhodes} J.,  {Refregier} A.,    {Groth} E.~J.,  2000, ApJ, 536, 79

\bibitem[\protect\citeauthoryear{Semboloni, Mellier, Waerbeke, Hoekstra,
  Tereno, Benabed, Gwyn, Fu, Hudson, Maoli \& Parker}{Semboloni
  et~al.}{2006}]{CFHTLS_deep}
Semboloni E.,  Mellier Y.,  Waerbeke L.~V.,  Hoekstra H.,  Tereno I.,  Benabed
  K.,  Gwyn S.,  Fu L.,  Hudson M.~J.,  Maoli R.,    Parker L.,  2006, A\&A,
  452, 51

\bibitem[\protect\citeauthoryear{{Spergel et al.}}{{Spergel et
  al.}}{2006}]{WMAP06}
{Spergel et al.} D.,  2006, ApJ, submitted, astroph/0603449

\bibitem[\protect\citeauthoryear{{Takada} \& {White}}{{Takada} \&
  {White}}{2004}]{TakadaWhite}
{Takada} M.,  {White} M.,  2004, ApJL, 601, L1

\bibitem[\protect\citeauthoryear{{Trujillo}, {Carretero} \&
  {Patiri}}{{Trujillo} et~al.}{2006}]{Trujillo}
{Trujillo} I.,  {Carretero} C.,    {Patiri} S.~G.,  2006, ApJL, 640, L111

\bibitem[\protect\citeauthoryear{{Vale} \& {White}}{{Vale} \&
  {White}}{2003}]{ValeWhite}
{Vale} C.,  {White} M.,  2003, ApJ, 592, 699

\bibitem[\protect\citeauthoryear{{van den Bosch}, {Abel}, {Croft}, {Hernquist}
  \& {White}}{{van den Bosch} et~al.}{2002}]{vdBosch02}
{van den Bosch} F.~C.,  {Abel} T.,  {Croft} R.~A.~C.,  {Hernquist} L.,
  {White} S.~D.~M.,  2002, ApJ, 576, 21

\bibitem[\protect\citeauthoryear{{Van Waerbeke}, {Mellier} \& {Hoekstra}}{{Van
  Waerbeke} et~al.}{2005}]{vWb04}
{Van Waerbeke} L.,  {Mellier} Y.,    {Hoekstra} H.,  2005, A\&A, 429, 75

\bibitem[\protect\citeauthoryear{{Van Waerbeke}, Mellier, Radovich, Bertin,
  Dantel-Fort, McCracken, F\`evre, Foucaud, Cuillandre, Erben, Jain, Schneider,
  Bernardeau \& Fort}{{Van Waerbeke} et~al.}{2001}]{vWb01}
{Van Waerbeke} L.,  Mellier Y.,  Radovich M.,  Bertin E.,  Dantel-Fort M.,
  McCracken H.,  F\`evre O.~L.,  Foucaud S.,  Cuillandre J.,  Erben T.,  Jain
  B.,  Schneider P.,  Bernardeau F.,    Fort B.,  2001, A\&A, 374, 757

\bibitem[\protect\citeauthoryear{{Van Waerbeke}, White, Hoekstra \&
  Heymans}{{Van Waerbeke} et~al.}{2006}]{VW06}
{Van Waerbeke} L.,  White M.,  Hoekstra H.,    Heymans C.,  2006, Astroparticle
  Physics accepted astroph/0603696

\bibitem[\protect\citeauthoryear{{White}}{{White}}{2002}]{TreePM}
{White} M.,  2002, ApJS, 143, 241

\bibitem[\protect\citeauthoryear{Wolf, Meisenheimer, Rix, Borch, Dye \&
  Kleinheinrich}{Wolf et~al.}{2003}]{Wolf03}
Wolf C.,  Meisenheimer K.,  Rix H.-W.,  Borch A.,  Dye S.,    Kleinheinrich M.,
   2003, A\&A, 401, 73

\bibitem[\protect\citeauthoryear{{Yang}, {van den Bosch}, {Mo}, {Mao}, {Kang},
  {Weinmann}, {Guo} \& {Jing}}{{Yang} et~al.}{2006}]{Yang06}
{Yang} X.,  {van den Bosch} F.~C.,  {Mo} H.~J.,  {Mao} S.,  {Kang} X.,
  {Weinmann} S.~M.,  {Guo} Y.,    {Jing} Y.~P.,  2006, MNRAS, pp 528--+

\bibitem[\protect\citeauthoryear{{Zehavi et al.}}{{Zehavi et
  al.}}{2004}]{zehavi04}
{Zehavi et al.} I.,  2004, ApJ, 608, 16

\end{thebibliography}
\label{lastpage}

\end{document}